\documentclass[
    aps,
    prd,
    reprint,
    superscriptaddress,
    amsmath,
    amssymb,
    floatfix
]{revtex4-2}
\pdfoutput=1 
\usepackage{amsmath}
\usepackage{graphicx}

\usepackage{xspace}
\usepackage{color}
\usepackage{xcolor}
\usepackage{rotating}
\definecolor{light-gray}{gray}{0.8}

\newcommand{\un}[1]{\ensuremath{\,\rm #1}}

\newcommand{\GeV}{\ensuremath{\rm GeV}\xspace}
\newcommand{\TeV}{\ensuremath{\rm TeV}\xspace}

\newcommand{\ZZero}{\ensuremath{{\rm Z^0}}\xspace}

\newcommand{\cmnt}[1]{}

\newcommand{\pt}{\ensuremath{p_{\mathrm{T}}}}
\newcommand{\MeVc}{\ensuremath{\mathrm{MeV}/c}}
\newcommand{\GeVc}{\ensuremath{\mathrm{GeV}/c}}

\newcommand{\drvd}{{\rm d}}

\begin{document} 

\title{Anomalous soft photon production can be understood as photon emission during hadronisation}

\author{Martin V\"olkl}
\email{m.a.volkl@bham.ac.uk}
\affiliation{School of Physics and Astronomy, University of Birmingham,
Edgbaston, Birmingham, B15 2TT, UK}

\author{Klaus Reygers}
\email{reygers@physi.uni-heidelberg.de}
\affiliation{Physikalisches Institut, Ruprecht-Karls-Universit\"at Heidelberg,
Im Neuenheimer Feld 226, Heidelberg, Germany}

\begin{abstract}
Several experiments reported measurements of soft photons in hadronic collisions far above the predictions from QED. This work investigates possible production during the hadronisation process. It is shown that the assumption of isotropic emission of photons during the hadronisation process leads to a consistent description of the experimental data.
\end{abstract}

\maketitle
\flushbottom

\section{Introduction}
\label{sec:intro}

Since the 1980s, a number of experiments measured an excess of soft photon production in hadronic collisions beyond the expectation from infrared divergence. Two of the most recent results by the WA102 \cite{WA102SoftPhotons} and DELPHI \cite{DELPHI:2005yew,DELPHI:2010cit} collaborations, showed an anomalous production about a factor of four above the expectation. DELPHI measured the dependence of the production with a variety of jet properties, with a particularly unexpected result being the strong scaling with the number of neutral particles in the jet. A further measurement by the DELPHI Collaboration in $e^+ e^- \rightarrow \mu^+\mu^-$ collisions \cite{DELPHI:2007nmh} found results consistent with the expectations from soft divergence.

A number of possible explanations have been put forward \cite{Bailhache:2024mck}. The objective of this work is to examine, whether the excess of soft photons can be consistently explained by photon emission during the hadronisation process. For this purpose, the parameters of the model will be fixed using the WA102 measurement of the photon \pt-distribution. The result will then be compared to the DELPHI measurements of the photon \pt- and $\theta$-distributions as well as the scaling with different jet properties.

\section{Photon production processes}
\label{sec:Production}

\subsection{Coherent production}
\label{sec:LowProduction}

Infrared divergences of cross section appear in calculations in quantum field theory for both vertex corrections and photon emissions. Based on the work of Bloch and Nordsieck~\cite{Bloch:1937pw} and a more modern treatment by Weinberg~\cite{Weinberg:1965nx}, the divergences cancel out leading to finite cross sections. The cancellation is motivated by the fact, that detectors cannot measure arbitrarily low photon energies. The calculation of photon emission in the soft limit contains a factorisation of the photon emission process and the hard scattering process:

\begin{align}
    \frac{{\rm d}N^{\gamma}}{{\rm d}^3\vec{k}} =& \frac{\alpha}{(2\pi)^2}\frac{-1}{E_{\gamma}}\int\left( {\rm d}^3 \vec{p}_1 \ldots {\rm d}^3 \vec{p}_{N_f} \right) \nonumber \\
    & \times \left( \sum_{{\rm Particle}~i} \frac{\eta_i e_i \bf{p}_i}{{\bf{p}_i} {\bf{k}}} \right)^2 \frac{{\rm d}N^{\rm H}}{{\rm d}^3 \vec{p}_1 \ldots {\rm d}^3 \vec{p}_{N_f}} ~. \label{eq:LowFormula}
\end{align}

The sum contains all incoming and the $N_f$ outgoing particles: $\vec{p}_i$ and $\bf{p_i}$ are the particles' three- and four-momenta, while $k$ is that of the photon. $e_i$ are the particles' charges and $\eta_i$ is a prefactor that is $\pm 1$ for incoming and outgoing particles respectively. Each term in the sum corresponds to a Feynman diagram with a single photon emitted from the corresponding particle. The sum thus corresponds to a coherent production from all these contributions at once. Based on Francis Low's work~\cite{Low:1958sn}, where he also calculated the next order contribution, this is often referred to as the Low formula.

This can also be used to calculate the soft photon production from a single event, e.g.:

\begin{align}
    \label{eq:LowFormula2}
    \frac{d^3{\rm N}}{d E_\gamma d \vartheta d \phi} =& \frac{-\alpha}{(2 \pi)^2} E_{\gamma} \sin \vartheta 
     \left( \sum_{{\rm Particle}~i} \frac{\eta_i e_i {\bf{p}_i}}{{\bf{p}_i} {\bf{k}}} \right)^2  \sim \frac{1}{E_{\gamma}} ~, 
\end{align}

where $\vartheta$ is the polar angle. Within a particular solid angle, the photon energies will follow a $1/E_\gamma$ distribution. This will thus be the dominant process in the limit of small photon energies. The formula is valid only if (i) photon emissions from internal lines are negligible and (ii) the energy lost to the emitted photon does not substantially change the cross section of the hadronic interaction. Low~\cite{Low:1958sn} suggests that this implies $E_\gamma / E_{\rm COM} \ll 1$ for the $2\!\rightarrow\!2$ scattering processes that he considered. Given that the production is a coherent sum of several contributions connected to incoming and outgoing particles, this also implies that the timescale of the process has to be short and the extent of the system small enough: $E_\gamma \Delta t \ll 1$ and $E_\gamma \Delta x \ll 1$ with $\Delta t$ and $\Delta x$ being typical scales of the system~\cite{Bailhache:2024mck}.

This radiation is referred to as soft photon radiation, inner or internal bremsstrahlung, or Low photons. Here, we will refer to it as inner bremsstrahlung and use the term soft photons for low-energy photons from any source.

\subsection{Incoherent production}
\label{sec:IncoherentProduction}

One interesting question is to ask what would happen if the emissions were not coherent and if this would explain an excess in photon radiation. Here, it is important to remember that the individual contributions in the sum in eq.\ \ref{eq:LowFormula} do not represent processes. Instead they originate in Feynman diagrams with an additional outgoing photon line from this charged particle. To remedy this, one can consider processes, where every incoming particle is first slowed down to rest, and every outgoing particle is accelerated from rest. This means that for every particle, two terms contribute: the original $(\eta_i e_i \bf{p}_i)/({\bf{p}_i} {\bf{k}})$ and $(-\eta_i e_i \bf{p}_{i,0})/({\bf{p}_{i,0}} {\bf{k}})$ with ${\bf{p}_{i,0}}=(m_i, \vec{0})$. So the sum in eq.\ \ref{eq:LowFormula} can be rewritten as:

\begin{equation}
    \left( \sum_{{\rm Particle}~i} \frac{\eta_i e_i \bf{p}_i}{{\bf{p}_i} {\bf{k}}} \right)^2 = \left( \sum_{{\rm Particle}~i} \frac{\eta_i e_i \bf{p}_i}{{\bf{p}_i} {\bf{k}}} - \frac{\eta_i e_i \bf{p}_{i,0}}{{\bf{p}_{i,0}} {\bf{k}}} \right)^2 ~.
\end{equation}

This is equal since the particle mass cancels out in the second term and all contributions $\eta_i e_i$ have to add up to zero due to charge conservation. In this view, the collision contains several processes: All incoming particles are first slowed down to rest and all outgoing particles are accelerated from rest, with all the soft photon emission from all processes contributing coherently. If these processes were to happen at different times or positions, then for a given direction and wavelength, this would introduce a phase shift:

\begin{align}
    &\left( \sum_{{\rm Particle}~i} \frac{\eta_i e_i \bf{p}_i}{{\bf{p}_i} {\bf{k}}} - \frac{\eta_i e_i \bf{p}_{i,0}}{{\bf{p}_{i,0}} {\bf{k}}} \right)^2 \nonumber \\
    &\rightarrow -\left| \left( \sum_{{\rm Particle}~i} \left( \frac{\eta_i e_i \bf{p}_i}{{\bf{p}_i} {\bf{k}}} - \frac{\eta_i e_i \bf{p}_{i,0}}{{\bf{p}_{i,0}} {\bf{k}}} \right) e^{i\phi_i} \right)^2 \right| ~,
\end{align}

with $\phi_i$ the phase for the process involving particle $i$. The phase difference for the incoming particles should be $0$. Due to the change in phases with wavelength and direction, this is not yet useful. But in a very complex system, the phases could be large and thus essentially random. The typical measured photon energies are of the order of $0.2\text{--}1\,\GeV$, so the system size would only have to be similar to a proton. In this limiting case, one would sum over all possible combinations of phases, which would lead to a sum of the squares of the individual contributions:

\begin{align}
    &\idotsint \drvd \phi_1 \cdots \phi_n \left| \left( \sum_{{\rm Particle}~i} \left( \frac{\eta_i e_i \bf{p}_i}{{\bf{p}_i} {\bf{k}}} - \frac{\eta_i e_i \bf{p}_{i,0}}{{\bf{p}_{i,0}} {\bf{k}}} \right) e^{i\phi_i} \right)^2 \right| \nonumber \\
    &= -\sum_{{\rm Particle}~i}  \left( \frac{\eta_i e_i \bf{p}_i}{{\bf{p}_i} {\bf{k}}} - \frac{\eta_i e_i \bf{p}_{i,0}}{{\bf{p}_{i,0}} {\bf{k}}} \right)^2 .
\end{align}

This corresponds to a fully decoherent production of photons: all processes radiate independently. One picture for this would be that some common, intermediate state is created --- like a QGP --- where the charges are mostly at rest and then the outgoing particles are created from this system at different times and positions. Given that the charges are at rest in one particular frame - the COM frame - this model breaks the Lorentz invariance of the original formula. The full stopping and acceleration of all charges is quite a strong assumption about the system. Figure~\ref{fig:WA102} shows that this decoherence would indeed increase the soft photon yield while still giving a $1/E_\gamma$ \pt-dependence. This can be seen by the comparison of the orange and blue lines. However, the increase is only of the order of a factor of 1.5, less than the factor 4 observed in the measurements. Thus, the photons in the measurements are most likely not from inner bremsstrahlung and other processes need to be considered.

\subsection{Photons from hadronisation}

The experimental observation of the increased soft photon production mostly concern hadronic collisions. In particular, a measurement by DELPHI in $e^+ e^-\!\rightarrow\!\mu^+\mu^-$ found no significant excess beyond the expectation from the soft divergence~\cite{DELPHI:2007nmh}. This suggests that the photons might be connected to QCD effects. The explanation explored in this work is that they are connected to the nonperturbative process of hadronisation.

The basic setup of the model is very simple: Each hadron produced in a sufficiently complex hadronic collision has a chance of radiating a photon, which is then emitted isotropically. The energy distribution of the emitted photons is assumed to be the same for all hadrons and systems. The main part of this work will be answering the question whether this leads to a consistent description for the photon excess measurements. In the following, we will attempt to justify this model.

One extreme case of hadronisation is the freeze-out of the QGP in heavy-ion collisions. Measurements indicate that even for the production of a quark-gluon plasma, with deconfined charges, hadronisation remains a local process. This can be seen from the success of the statistical hadronisation models for the overall particle yields~\cite{Andronic:2017pug} and the success of the Cooper-Frye freeze-out in hydrodynamic descriptions~\cite{SHEN201661,ALICE:2018yph}. In both cases, the hadronisation description happens locally, without any correlations with light particle production elsewhere in the system. So even with deconfinement, the production of colour-neutral objects only seems to involve a local region. This could be facilitated by assuming that before hadronisation some colour-neutral pre-hadrons form which then become the final state hadrons. These are objects which still interact extensively with the rest of the system. As the interaction ceases, hadrons start to appear.

While the pre-hadrons still interact with the medium, their masses may not be the mass of the particle in a vacuum. An analogy for this effect would be the Stark and Zeeman effects of atomic physics: A background field changes the energy states. In QCD, chiral restoration is an example of strong fields changing the hadron masses. In the hadronic processes investigated in anomalous soft photon production, the background field is created by the rest of the collision system. The hadronisation process then corresponds to a decrease of the background fields towards zero and an approach of the hadron mass to its vacuum value. To reach the final mass, the pre-hadron has to either gain or lose energy from the rest of the system. In the electromagnetic case, this would happen by the exchange of photons. However, the same process cannot happen via the exchange of a single gluon, because this would create a colour charge of the state. While more complex QCD based exchanges may still happen, these are suppressed. Thus, the emission of a photon may be a competitive process for reaching the vacuum mass.

For other hadronic processes, similar considerations apply. In jet fragmentation, colour neutral objects are created at the end of the string breaking process. But these are still near other fragments and strong fields may exist. In general, hadrons start to exist as the interactions with the rest of the system cease. This means that the production of hadrons can be considered to involve similar field strengths during hadronisation in different systems. Thus, we will assume in the following that the emission of the soft photons by these hadrons will follow the same law in all systems, as long as energies and multiplicities are sufficiently high. Previously, several publications have suggested photon emission in the hadronisation phase~\cite{PhysRevC.106.034906, PhysRevC.107.024916}. In these cases, the emission was assumed for the specific mechanism of quark coalescence and in the context of QGP freezeout in heavy ion collisions. For the model presented here, photon emission at hadronisation is expected to be present in a large number of systems and independent of whether the process happens via coalescence or in another way. Another model with intermediate states is the Glob model by Lichard and Van Hove~\cite{LICHARD1990605,Lichard:1994qt}. There, the authors assume an intermediate, long lived state of cold quark-gluon plasma, which can radiate photons via processes such as $q\bar{q}\rightarrow g\gamma$. In contrast to this, the model discussed here associates the photon production with individual, final state particles.

The model has two ingredients: A probability for the emission of a photon and a distribution of the energy of the photons in the frame of the pre-hadron. If the mass of the pre-hadron is lower than that of the final hadron, the process must happen via the strong interaction. Thus, the probability should be $<\!0.5$ as long as only single photon emission is considered. If the mass is much higher than the final one, then the vacuum mass can be reached more effectively via the emission of a pion. This means that the photon emission energy must be lower than the pion mass.

Due to the complexity of the processes, there is no obvious choice for the form of the energy distribution, so the main aim here is to keep the description simple. In the following the photon energy distribution in the system of the pre-hadron will be represented by a power law, cut at the pion mass:
\begin{equation*}
    \frac{\drvd p}{\drvd E} = \begin{cases}
         \frac{1-\alpha}{m_\pi} \left(\frac{E}{m_\pi}\right)^{-\alpha} & {\rm for}~ E<m_\pi \\
        0 & {\rm else}
    \end{cases}
    ~.
\end{equation*}
The estimates shown later are created from Pythia 8.315 \cite{Bierlich:2022pfr} events, created using the Monash 2013 tune \cite{Skands:2014pea}. To keep the calculation simple, we have neglected the recoil of the photon on the particle. So to calculate the photon emission, a photon is sampled from the energy distribution, given a random direction, and then boosted according to the velocity of the emitting hadron. The two parameters, power law coefficient and emission probability, are fixed based on the WA102 measurement. This model is then applied to the DELPHI measurements to investigate whether this results in a consistent description.

\section{Comparison with measurements}

To test the model, it is compared with data taken at WA102 and DELPHI \cite{DELPHI:2010cit}. In these experiments, photons in the energy range $0.2\text{--}1\,\GeVc$ were measured using a variety of selection criteria and plotted against different variables. In all cases, the results were compared with an estimate of the photon production from inner bremsstrahlung. To compare to the results, it is important to include the specific selection criteria as well as to incorporate some of the detector properties. For this comparison, both the photons from the hadronisation production as well as the expectation from inner bremsstrahlung were calculated based on the expectation from Pythia 8. The soft bremsstrahlung expectation was then compared to the one from the publications to make sure that the modern event generator gives similar results and that all relevant detector and reconstruction effects were included. With these checks, the WA102 result was used to constrain the parameters of the model and the model using these parameters was then compared to the results from the DELPHI publications.

\subsection{WA102}

\begin{figure}[tbp]
\includegraphics[width=1.\columnwidth]{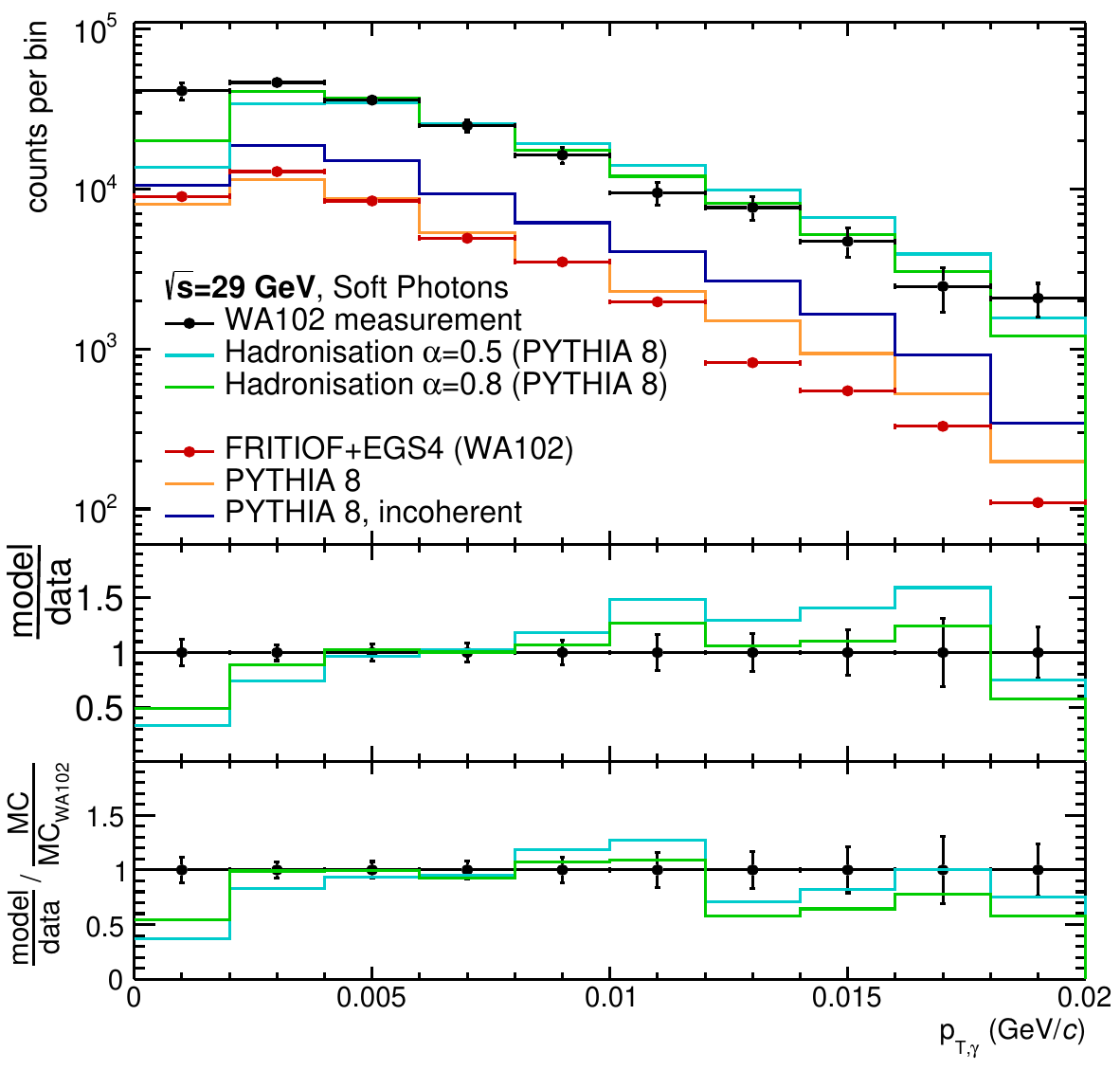}
\caption{\label{fig:WA102} Comparison of model with the efficiency corrected WA102 measurements for $\theta<20\un{mrad}$. The error bars are statistical.}
\end{figure}

The WA102 measurement \cite{WA102SoftPhotons} was taken in fixed target pp collisions with a $450\,\GeVc$ beam. Measurements were taken as a function of \pt\,in ranges of the azimuthal angle $\theta$, both of which were calculated relative to the beam axis and direction. The setup included a magnetic field of $B=1.2\,{\rm T}$. The photons were measured via the conversion on a lead sheet by the MWPCs behind it. To exclude contributions from regular bremsstrahlung from the lead sheet, an isolation cut was included: Photons needed to have at least a $3\un{mm}$ distance to the nearest charged track on the lead sheet or they were discarded. The hydrogen target had a length of $60\un{cm}$ and a distance of its centre to the lead sheet of $66\un{cm}$, meaning that the corresponding angle of the isolation cut varied from event to event. Only events with fewer than 8 charged tracks were considered for the analysis. Most of the excess soft photons were found at an azimuthal angle $\theta<20\un{mrad}$, so that is the result considered in this work.

\subsubsection{Cross-check using the estimation of inner bremsstrahlung}

In the WA102 publication \cite{WA102SoftPhotons}, the expected contribution from inner bremsstrahlung was estimated using the FRITIOF event generator together with a collaboration-created particle transport code. To reproduce this, the distance of the collision to the target was randomly varied between $36$ and $96\un{cm}$. The isolation cut was applied including bending of the charged tracks in the magnetic field. The WA102 points were extracted from the publication via a digitisation tool. The comparison shown in Figure~\ref{fig:WA102} is to the efficiency corrected result. Comparing the red and orange line, the expectation from inner bremsstrahlung is well reproduced by Pythia 8. The agreement suggests that the most relevant selection criteria are included. At the higher \pt\,end, Pythia 8 somewhat overestimates the expected production. It is not clear what causes this discrepancy. It could either be a difference in the event generators, or a detector effect missing in the Pythia 8 based analysis.

\subsubsection{Extracting the model parameters}

The reasonable reproduction of the soft photon expectation suggests that this framework can now be used to test the model of additional photon production during hadronisation. For this purpose, a power law exponent of $-0.8$ yields a \pt-distribution in good agreement with the measurement, apart from the lowest \pt-interval, as shown in Figure~\ref{fig:WA102}. The difference in the lowest \pt-interval may be due to a discrepancy with the model or due to imperfect implementation of the detector effects. Thus, the photon emission probability was estimated with and without this interval, yielding $0.134$ and $0.119$ respectively.

Since the PYTHIA 8 calculation without full detector effects seems to create a higher yield for inner bremsstrahlung than the WA102 results at high \pt, this might point to a difference in the way the detector is treated. To include such effects, the ratio of model and data is also corrected for the ratio of these estimates. In this case, a power law coefficient of $-0.5$ leads to a mostly flat ratio. This does not have a substantial effect on the total yield. The emission probabilities corresponding to the pure yield with and without the first bin are $0.136$ and $0.117$ respectively. The statistical uncertainties of the measurement are so small that they are not considered in the parameters. The difference in the power law coefficient turns out to be much more relevant.

To implement a function that gives additional contributions in the first bin only, more contributions would be needed at a low energy in the pre-hadrons system. However, these photons need to be emitted by a very fast particle to reach the lower energy limit of the measurement. This means that they will be suppressed by the isolation cut as they hit the detector near a high energy particle.

Thus, the model is able to describe the WA102 result fairly well, albeit with the use of two free parameters: the power law coefficient and the emission probability, with the latter acting as an amplitude prefactor. To test the model more thoroughly, these parameters are now applied to the DELPHI measurements, which considered photons in the context of jets from Z boson decay. The power law coefficients represent reasonable boundaries between which the true function may lie according to the WA102 measurement.

\subsection{DELPHI measurements relative to the jet axis}
\label{sec:DELPHI1}

\begin{figure}[tbp]
\includegraphics[width=1.\columnwidth]{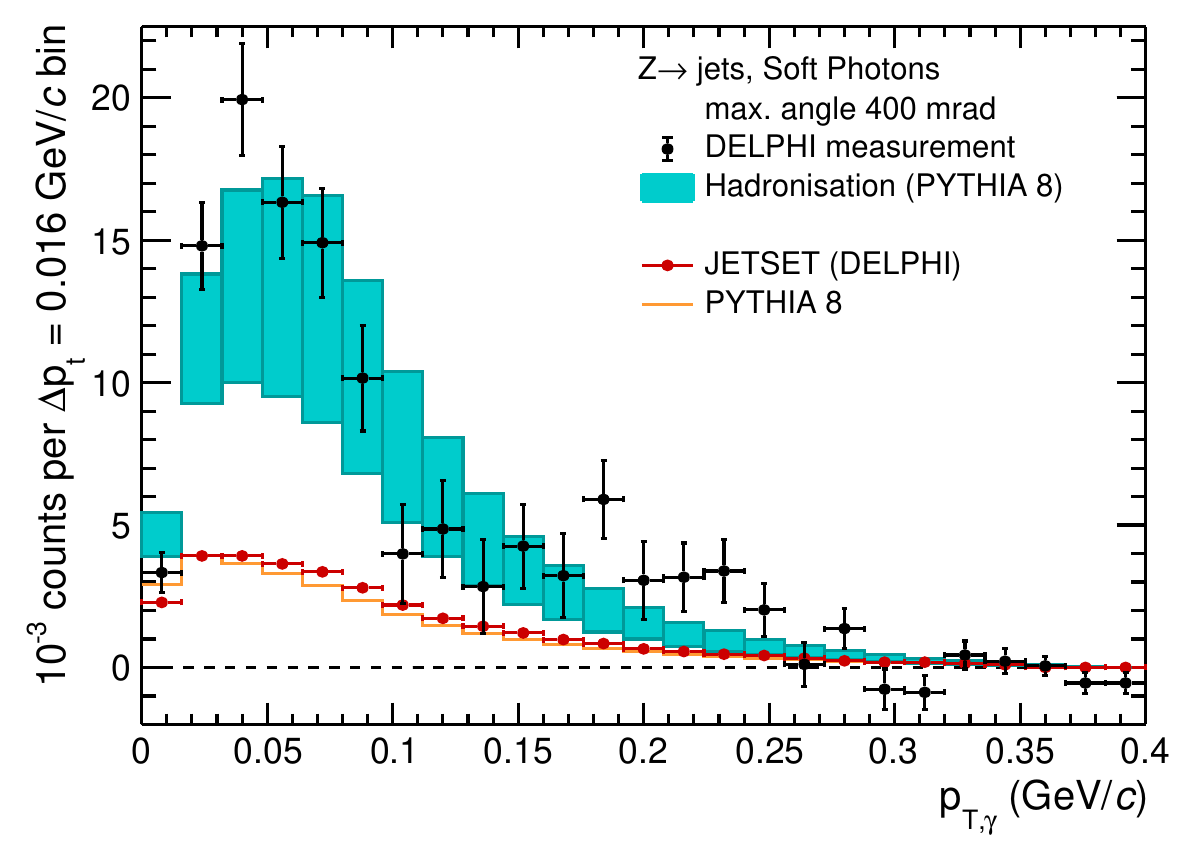}
\hfill
\includegraphics[width=1.\columnwidth]{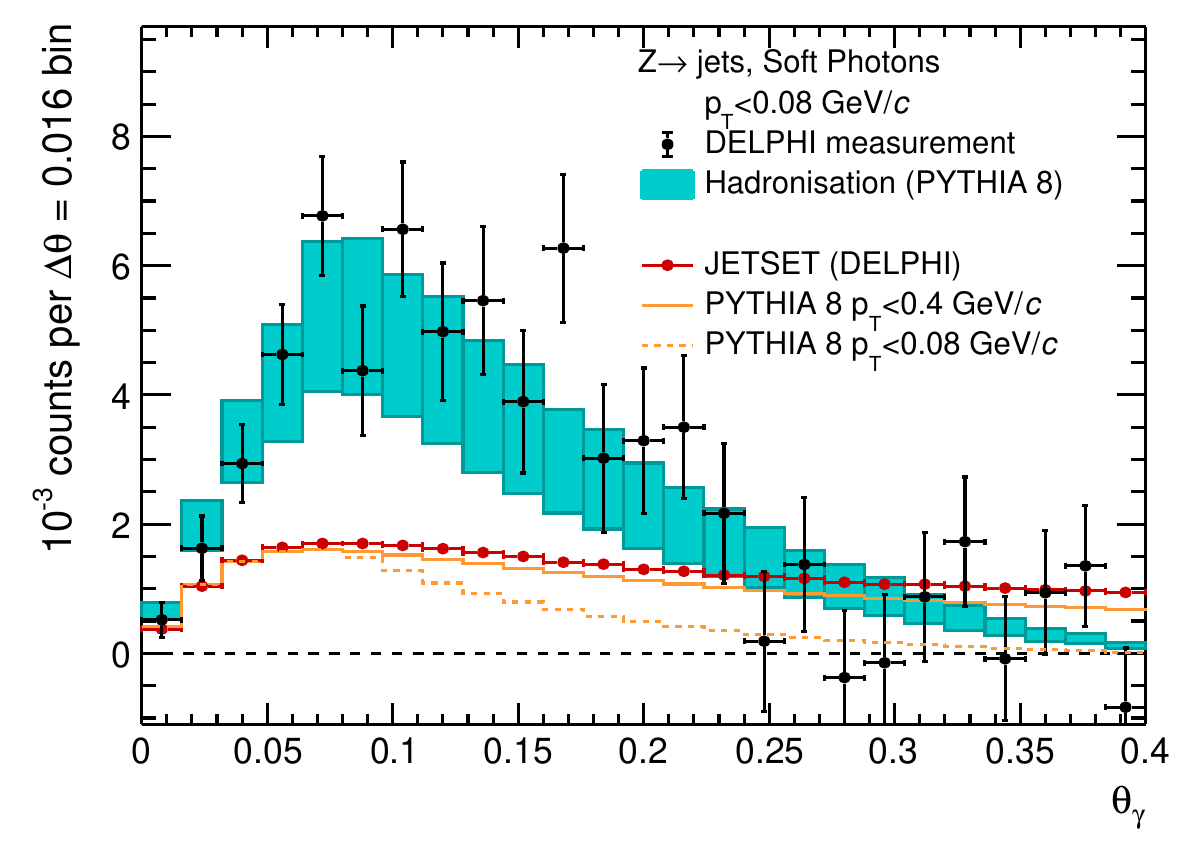}
\caption{\label{fig:DELPHI1} Comparison of model with DELPHI measurements.}
\end{figure}

The earlier of the publications by the DELPHI Collaboration \cite{DELPHI:2005yew} measured soft photons at LEP in the decays of \ZZero bosons, $e^+ e^- \rightarrow \ZZero \rightarrow {\rm hadrons}$. Differently from WA102, the transverse momentum and polar angle in this measurement are defined relative to the jet axis of the jet associated with this particle. In both DELPHI publications, the jet reconstruction was done by LUCLUS \cite{Sjostrand:1982am} with a resolution parameter of $d_{\rm join}= 3\,\GeVc$. To approximate this, ClusterJet was used within the Pythia 8 framework, using a $y$-scale parameter of $0.01$ with the JADE \cite{BETHKE1988235} distance measure. This setup was also used as a cross-check by the DELPHI Collaboration and performed similarly. Unlike WA102, DELPHI did not use an isolation cut. To reduce the background from $\ZZero \rightarrow \tau \tau$, the measurement required at least two jets, five charged tracks, and high visible energy of $E_{\rm vis} > 0.2 E_{\rm COM}$. This was reproduced by adding up the number and energy of the final state charged particles in the Pythia 8 event. In the laboratory system, only jets with an axis polar angle between $30$ and $150^\circ$ were considered. The measurement considered charged tracks with $p>400\,\MeVc$ and a polar angle to the beam axis between $20$ and $160^\circ$. Within the jets there were two kinds of electron veto: Firstly, jets with identified electrons were removed. This was reproduced by ignoring all jets with electrons for the modelling. Secondly, for jets with a single charged particle, that particle must be identified to not be an electron. As for this case the particle will usually have a very high momentum, this was reproduced by removing all jets with a single charged particle from consideration.

\subsubsection{Cross-check using the estimation of inner bremsstrahlung}

The main results of the measurement are the $\theta$, \pt, and $\pt^2\,$ distributions of soft photons after subtraction of a substantial decay photon background. These variables were calculated with respect to the axis of the jet closest to the photon. Here, we will compare to the efficiency corrected results in $\theta$ and \pt. DELPHI measured $\theta$ up to 0.4 and \pt\,up to $0.4\,\GeVc$. The angle of photons to the jet axis of the closest jet can reach about $\pi/2$ for a two jet event. The Pythia 8 reproduction of the inner bremsstrahlung results in \pt\,is close to the results from DELPHI, assuming that the plot only shows photons within $\theta<0.4$. The $\theta$ distribution was measured in the range $\pt<0.08\,\GeVc$. However, this seems not to be the case for the inner bremsstrahlung results in the same plots. For $E_\gamma>0.2\,\GeV$ and $\pt<0.08\,\GeVc$, $0.4$ represents the edge of the available phase space, so the distribution should go to zero, which it does not. Assuming that the limit for this calculation was $\pt<0.4\,\GeVc$, the reproduction with Pythia 8 agrees fairly well, slightly undershooting the JETSET result at high $\theta$ as shown in Figure~\ref{fig:DELPHI1} (bottom). Thus, the setup with Pythia 8 reproduces the major effects of the measurement and can be used to test the model.

\subsubsection{Comparison with the model}

This comparison is also shown in Figure~\ref{fig:DELPHI1}. The two power laws form the upper and lower limit of the model, with $\alpha=0.5$ corresponding to the upper edge of the area. While the peak in the \pt-distribution is slightly higher in the measurement, the model shows good agreement in both cases shown in Figure~\ref{fig:DELPHI1}. This shows that the model based on the parameters set by the WA102 results is able to qualitatively and quantitatively describe the DELPHI $\theta$ and \pt-distributions.

\subsection{DELPHI measurements as a function of the jet properties}
\label{sec:DELPHI2}

\begin{figure}[tbp]
\centering
\includegraphics[width=1.\columnwidth]{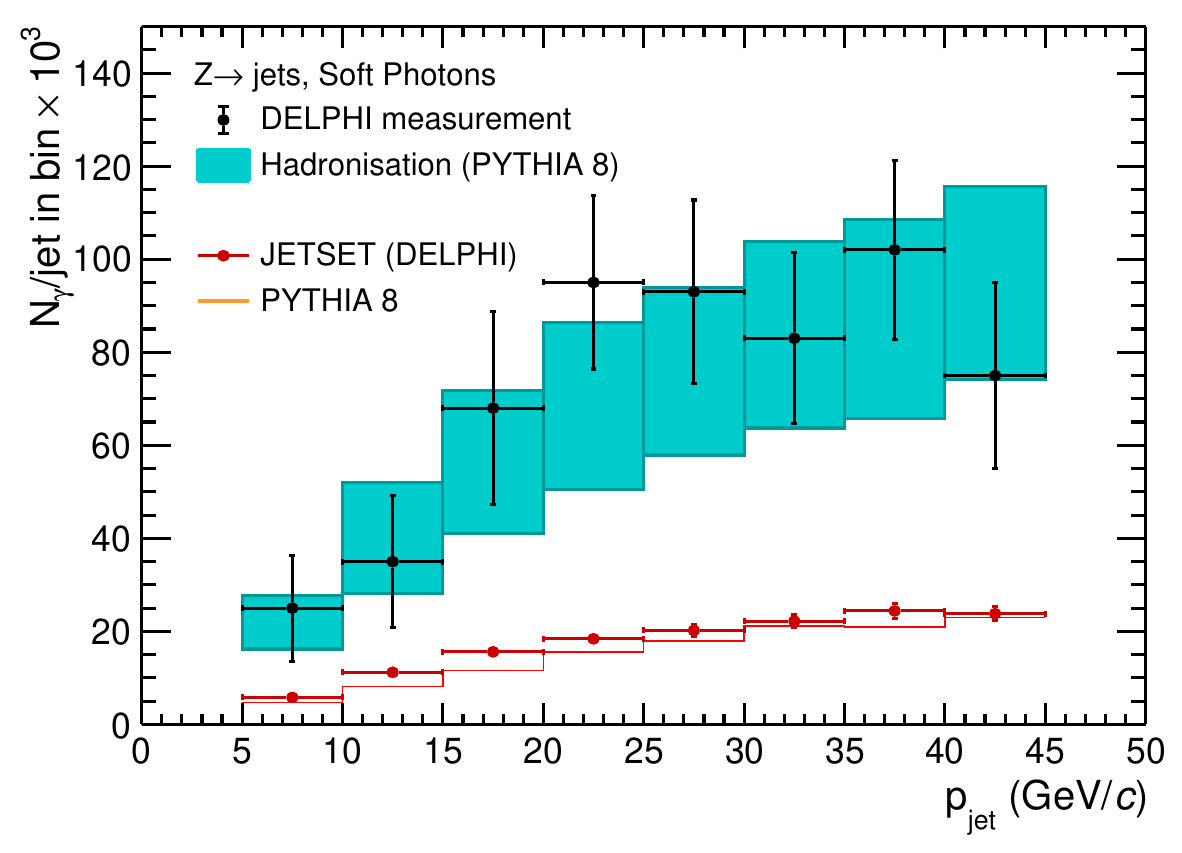}\\
\includegraphics[width=1.\columnwidth]{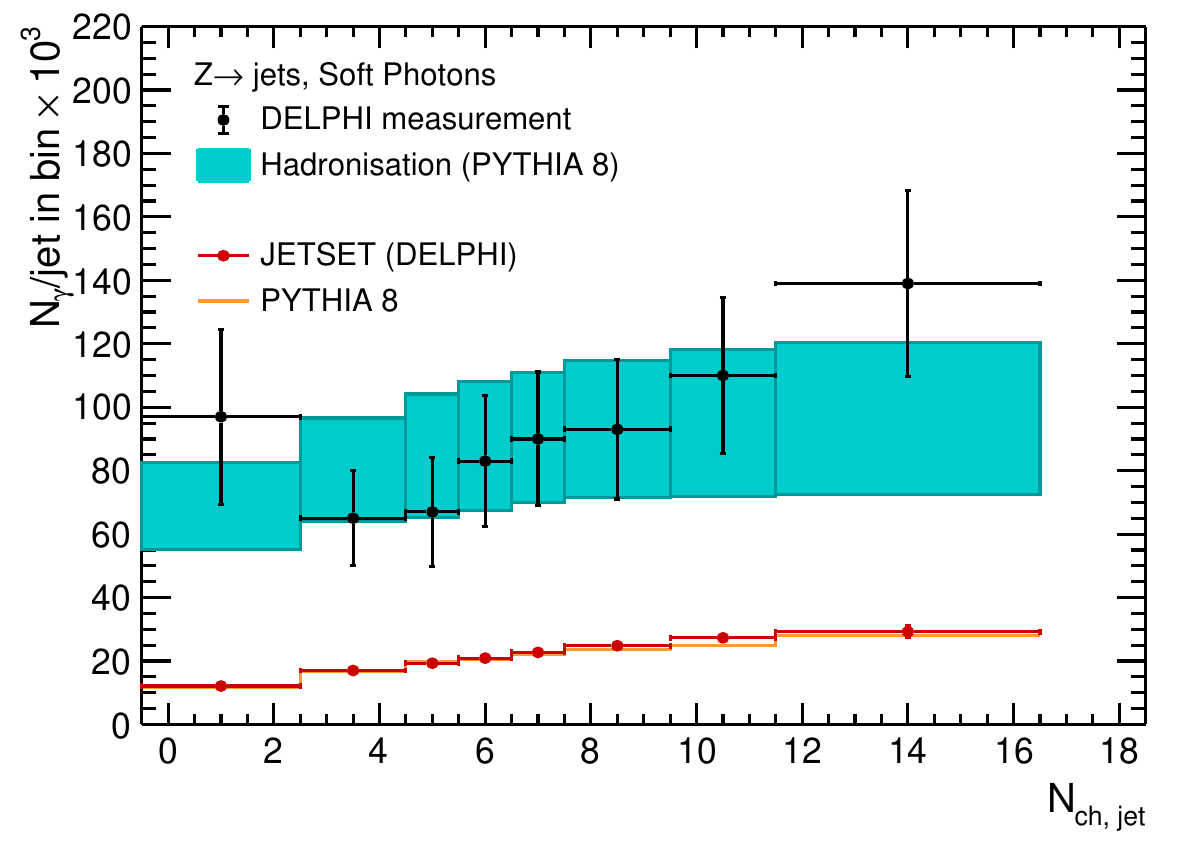}\\
\includegraphics[width=1.\columnwidth]{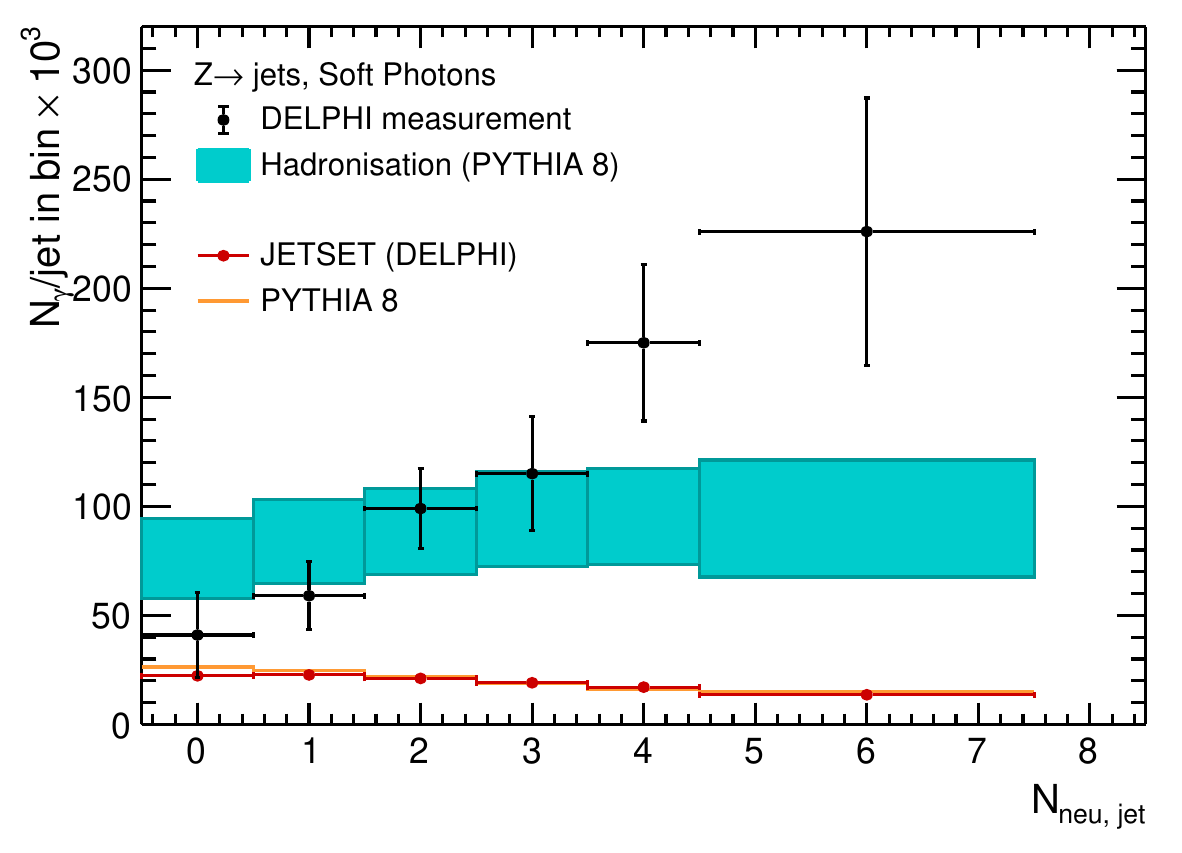}
\caption{\label{fig:DELPHIJetProperties} Comparison of model with DELPHI measurements as a function of jet properties~\cite{DELPHI:2010cit}.}
\end{figure}

In their more recent publication on the topic \cite{DELPHI:2010cit}, the DELPHI collaboration investigated the anomalous production in jets more thoroughly. This time without an angle or \pt-cut, photons in the energy range $0.2\text{--}1\,\GeVc$ were again associated with the closest jet. The scaling of the anomalous photon production with several variables was tested. One particularly surprising result was the strong scaling with the number of neutral particles in the jet, as these do not contribute in the Low formula. As before, the events were required to have at least five charged tracks and a visible energy of $E_{\rm vis} > 0.2 E_{\rm COM}$. As for the measurement in section \ref{sec:DELPHI1}, jets with identified electrons were vetoed, which was included by removing jets with electron constituents from consideration in the Pythia 8 comparisons.

Here, we will compare to the scaling of the soft photon yield as a function of the total jet momentum, the number of charged particles and the number of neutral particles in the jet. DELPHI uses both charged and neutral particles to reconstruct the jets. For the reproduction, the sum of the constituent momenta was taken as the jet momentum. The charged particle multiplicity was calculated from the charged particles reconstructed within the jet with $\pt>0.2\,\GeVc$. The neutral particle multiplicity was calculated by adding the number of photons in the jet with a weight of 0.5 if they had $E>1\,\GeV$ and the number of other neutral particles with a weight of 1 if their energy exceeded $2\,\GeV$.

\subsubsection{Cross-check using the estimation of inner bremsstrahlung}

Figure~\ref{fig:DELPHIJetProperties} shows that the inner bremsstrahlung expectations are reasonably well reproduced by the simplified calculations with Pythia 8. The production rises as a function of the momentum and number of charged particles and decreases with the number of neutral particles. This suggests that the setup with Pythia 8 captures the main effects of the measurement and can be used to test the model.

\subsubsection{Comparison with the model}

The model comparison is also shown in Figure~\ref{fig:DELPHIJetProperties}. The rise of the soft photon yield with the jet momentum is well reproduced by the model. Similarly, the increase with the number of charged particles is fully consistent with the measurement. The model shows a rise in the per jet photon yield with increasing number of neutral particles, which runs counter to the trend for the inner bremsstrahlung expectation. Here, it is useful to remember that the upper and lower limit correspond to the $\alpha=0.5$ and $\alpha=0.8$ energy distributions in the centre of momentum frame. So for each individual setting, there is a monotonic rise in the number of photons per jet. The rise is less steep in the model than in the measurement. Given the large uncertainties of the measurement and the fact that they are mostly due to the systematic contributions, the model is still fairly consistent with the measurement.

\subsection{Effect of the selection criteria}

\begin{figure}[tbp]
\centering
\includegraphics[width=1.\columnwidth]{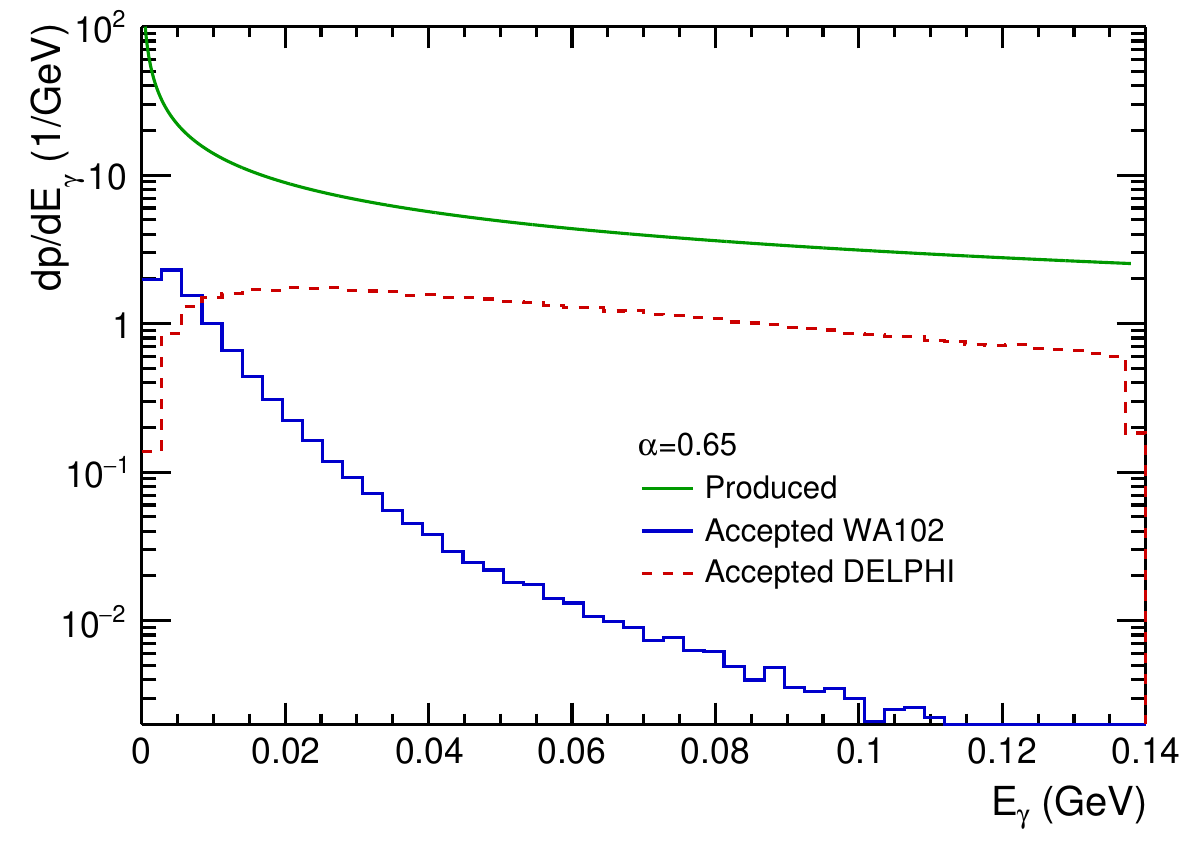}
\caption{\label{fig:SelectionEffect} Distribution of photon energies in the emitting particle rest frame with and without measurement selection. This corresponds to the \pt-distribution measurements of the experiments.}
\end{figure}

Due to the different collision type and analysis in the two experiments, they select different kinematic ranges of the photons. Figure~\ref{fig:SelectionEffect} shows the effect on photons that were emitted at different energies in the rest frame of the particle. The WA102 measurement selects preferentially photons from the lower energy end, while these are suppressed in the DELPHI case. This is due to the boosted system in the fixed-target setup of WA102 as well as the isolation cut.

\subsection{ALICE measurement of direct photons}

\begin{figure}[tbp]
\centering
\includegraphics[width=1.\columnwidth]{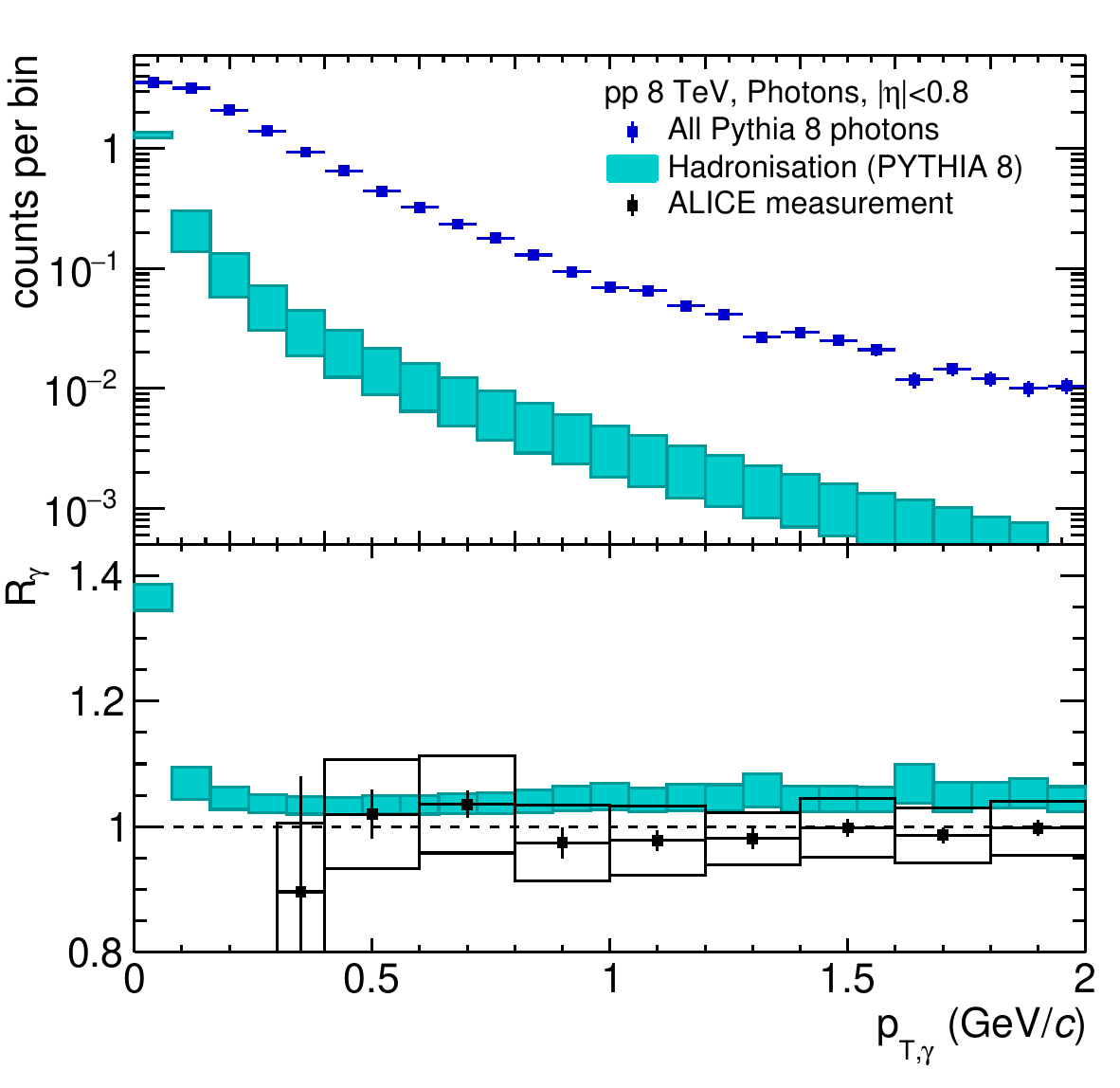}
\caption{\label{fig:ALICERgamma} The top plot shows the distribution of simulated decay photons and photons from hadronisation. The bottom plot shows the fraction of photons from hadronisation as well as the ALICE measurement~\cite{ALICE:2018mjj} of the fraction, $R_\gamma$.}
\end{figure}

The photons in this model are not necessarily at low energies. They can have substantial energies in the lab frame. This may allow additional tests of the model. Figure~\ref{fig:ALICERgamma} shows the production in $8\,\TeV$ pp collisions together with all photons produced by the Pythia 8 generator in this kinematic range. Rather than subtracting the background, ALICE published the quantity $R_\gamma$, which is the ratio of all measured photons to the contribution from hadronic decays only~\cite{ALICE:2018mjj}. Values above 1 quantify an excess photon production. Calculating this value with the model results yields values for $R_\gamma$ of about 1.05, which sits near the edge of the uncertainty of the ALICE measurement. Thus, with an additional decrease of the uncertainties, photons from hadronisation could be measured even in the \GeV-range, which would provide a useful test for the model.

\section{Conclusions}
\label{sec:Conclusions}

The anomalous production of soft photons in hadronic collisions is consistent with photon emission during the hadronisation process. A model based on isotropic emission of photons in the frame of the produced particles with a common energy distribution in this system can describe both the WA102 and DELPHI data consistently. This model also predicts an increase of photon production at higher momenta in particular at LHC energies. A modest improvement in measurement accuracy of photons in the \GeV-range would be sufficient to measure this in the future.

\appendix

\acknowledgments

We would like to thank J\"urgen Berges, Peter Braun-Munzinger, David Evans, Aleksas Mazeliauskas, and Johanna Stachel for many useful discussions. This work was supported in part by the Science and Technology Facilities Council (STFC), UK, and by the DFG (German Research Foundation) — Project No. 273811115 — SFB 1225 ISOQUANT.


\bibliography{bibliography.bib}

\end{document}